# Are there stars in Bluesky after the return of Donald Trump to the White House?

Wenceslao Arroyo-Machado[1]; Nicolas Robinson-Garcia[2,*]; Daniel Torres-Salinas[2]
warroyom@asu.edu; elrobin@ugr.es; torressalinas@ugr.es

[1] Center for Science, Technology and Environmental Policy Studies (CSTEPS), School of Public Affairs, Arizona State University, Phoenix, AZ 85004, USA
[2] Department of Information and Communication Sciences, University of Granada, Granada, Spain

*Corresponding author

**Abstract**

This study examines the shift in the scientific community from X (formerly Twitter) to Bluesky, its impact on scientific communication, and consequently on social metrics (altmetrics). We analysed 14,497 publications from multidisciplinary and Library and Information Science (LIS) journals between January 2024 and March 2025. The results reveal a notable increase in Bluesky activity for multidisciplinary journals in November 2024, likely influenced by political and platform changes, with mentions multiplying for journals like Nature and Science. In LIS, the adoption of Bluesky is different and shows marked variation between European and United States journals. Although Bluesky remains a minority platform compared to X over the whole period, when focusing on user engagement after the United States elections, we see a much more even distribution between the two platforms. In two LIS journals, Bluesky even surpasses X, while in most others, the difference in user engagement was no longer as pronounced, marking a significant change from previous patterns in altmetrics.

## 1. Introduction

On 5 November 2024, Donald Trump won the US presidency for a second term over his opponent Kamala Harris, among his closest supporters was Elon Musk, owner of the social media platform X (previously Twitter) after a controversial purchase in October 2022. A day after his triumph, Trump announced that Musk would be head of the Department of Government Efficiency under his administration. Therefore, over the next weeks, millions of users have moved to Bluesky (Holterman, 2024), a rival service also founded by Jack Dorsey, founder of Twitter. Bluesky essentially mimics the functionalities that Twitter had before Musk's take over. Among the users shifting, many of them seem to be researchers, pushing Altmetric to announce on 3 December 2024, that it will now also be tracking this new social media platform (Altmetric, 2024).

User migration across social media platforms is a common aspect of social media and affects the platforms where scientists share their results and interact with their peers (Jeong et al., 2023). Consequently, it also impacts the metric aspects of scholarly information, influencing how altmetrics are collected (e.g., Google Plus or Delicious). However, this is an interesting case due to two reasons. First, the difficulty to track social media discussion of publications has increased since Musk's takeover, pointing towards a strategy from Altmetric to redirect the traffic of mentions to scientific literature from X to Bluesky. Second, if that is the case, this move has important implications, as X, contrary to other deceased platforms, is the source representing the largest bulk of data offered by Altmetric.

The first objective of this brief communication is to verify whether there is evidence of a community shift from X to Bluesky, within the broader context of scholarly communication, from its inception to March 2025. Consequently, if such community shift is found, the second objective is to examine the differences in values and indicators between the two platforms. This study analyzes the coverage and number of mentions in two distinct sets of journals: three multidisciplinary journals and four Library and Information Science journals, comparing their presence on both X and Bluesky. This analysis will help us understand whether the activity reflected on Bluesky is comparable in volume to that on X, a crucial step in assessing the implications of incorporating Bluesky mentions and evaluating their significance or comparability to those from X. The results are expected to provide valuable insights into whether Bluesky currently serves as an effective platform for the dissemination of scientific information and the generation of altmetric indicators.

## 2. Methodology

We constructed a dataset of 14,497 publications from three multidisciplinary journals—*Nature*, *Science*, and *Proceedings of the National Academy of Sciences of the United States of America* (*PNAS*)—and four Library & Information Science (LIS) journals—*Journal of the Association for Information Science and Technology* (*JASIST*), *Journal of Informetrics*, *Scientometrics*, and *Quantitative Science Studies* (*QSS*)—from January 2024 to March 2025. This selection was made to ensure representation of both broad, multidisciplinary research and specialized fields, providing two distinct control groups to justify the robustness of our comparisons. We focus solely on this period as Bluesky opened the registration to its service in February 2024, when registration was invitation-only.

We used two complementary methods to retrieve mentions data. The first was a custom, ad hoc approach developed for this study, which involved extracting each record's DOI, resolving the corresponding URL, and collecting alternative associated URLs (e.g., open access versions or PDFs) provided by OpenAlex. These identifiers and URLs were then used to query the Bluesky API for mentions. We excluded title-based searches due to a high rate of false positives in top journals. This allowed us to obtain the total number of mentions on Bluesky directly. The second method relied on the Altmetric API, which was used to retrieve the number of unique users on X (formerly Twitter) who mentioned each research output. Since Altmetric also tracks Bluesky activity[1], we extracted from it the number of Bluesky users referencing each paper as well. However, as the free version of the Altmetric API does not provide the total number of mentions, only the number of users is available through this method. As a result, on the one hand, the total number of mentions on Bluesky from our custom method, and on the other, the number of users mentioning each document on both X and Bluesky via Altmetric. To ensure consistency in the comparison with X, we chose to rely on Altmetric's user counts for Bluesky as well. This also allowed us to identify and analyze potential discrepancies between the two methods. The Python script used to retrieve data from Bluesky and X via Altmetric is available on GitHub (https://github.com/Wences91/bluesky_altmetrics), and data exploration was carried out in R using descriptive statistics.

## 3. Results

Table 1 shows the coverage and average number of users mentioning papers for both platforms, distinguishing between two sources for Bluesky (ad-hoc method and Altmetric). For leading multidisciplinary journals, we observe that the percentage of papers with at least one mention on Bluesky is substantially lower than on X across all cases and sources. The only exception is *Nature*, where the mention rate using the ad-hoc method on Bluesky (92%) is comparable to that found on X (99%). To gain a more accurate understanding, it is essential to consider the number of unique users mentioning papers, where it becomes evident that the figures are notably lower for Bluesky, regardless of the data source. For instance, in *Nature*, the average number of accounts on X is 133.40 compared to 6.86 (ad-hoc) or 27.63 (Altmetric) on Bluesky. Focusing on Library and Information Science (LIS) journals, *Scientometrics*, *Journal of Informetrics*, and *QSS*, the percentage of papers mentioned on X ranges between 28% and 45%, while on Bluesky, the figures are significantly lower. For example, *Journal of Informetrics* achieves only between 5% and 8%, while *Scientometrics* reaches 16% and 11%, respectively. Furthermore, the drop in the number of user accounts mentioning papers is particularly notable on Bluesky, as seen in *QSS*, where the average falls from 26.93 on X to 0.60 and 12.44 on Bluesky.

---

[1] At the time of data collection, Altmetric only captures posts that include a direct link to the publication, which may lead to underrepresentation in some cases. Source: https://help.altmetric.com/support/solutions/articles/6000276192-bluesky

**Table 1.** *Coverage and average number of accounts mentioning papers by journal across X and Bluesky*

| Journal | Nr Papers | X Source: Altmetric | | Bluesky Source 1 (S1): Ad-hoc method Source 2 (S2): Altmetric | |
|---|---|---|---|---|---|
| | | Papers mentioned | Avg. Accounts | Papers mentioned | Avg. Accounts |
| *Nature* | 5836 | 5796 (99%) | 133.40 | S1 5380 (92%)<br>S2 2951 (51%) | 6.86<br>27.63 |
| *PNAS* | 5201 | 4236 (81%) | 18.68 | S1 1816 (35%)<br>S2 1569 (30%) | 0.96<br>3.64 |
| *Science* | 2662 | 2066 (78%) | 68.29 | S1 1305 (49%)<br>S2 1022 (38%) | 3.06<br>15.1 |
| *Scientometrics* | 400 | 178 (44%) | 5.92 | S1 64 (16%)<br>S2 44 (11%) | 0.34<br>0.48 |
| *Journal of Informetrics* | 143 | 40 (28%) | 1.31 | S1 7 (5%)<br>S2 12 (8%) | 0.06<br>0.25 |
| *JASIST* | 169 | 50 (30%) | 4.67 | S1 28 (17%)<br>S2 28 (17%) | 0.44<br>0.69 |
| *QSS* | 86 | 39 (45%) | 26.93 | S1 20 (23%)<br>S2 24 (28%) | 0.60<br>12.44 |
| ***Total*** | **14,497** | **12,405 (86%)** | **73.61** | **S1 8620 (59%)**<br>**S2 5650 (39%)** | **3.70**<br>**15.36** |

Although the overall data position Bluesky as a minor altmetric source, the evolution of mentions, potentially linked to political events, from January 2024 to March 2025 reveals some patterns (Figure 1). For multidisciplinary journals, we observe a clear surge in mentions beginning in September 2024 and continuing into early 2025, particularly around the United States presidential election and subsequent political events. For instance, *Nature* stands out, with mentions rising sharply from 440 per week in early November to 1,412 per week by mid-November. Notably, this rise in attention does not represent a mere burst in activity, rather, the volume of mentions stabilizes at a consistently high level in the months that follow. In contrast, the Library and Information Science (LIS) journals show much lower and more irregular levels of activity, but it is also important to note that their number of papers is markedly lower compared to the top multidisciplinary journals. However, all LIS journals experience a noticeable peak in mentions at the end of 2024. In both *Scientometrics* and *QSS*, this increase begins around November 2024 and is followed by a slight but prolonged rise in activity that extends into early 2025. However, even during these periods of increased attention, the absolute volume of mentions remains low, never exceeding 20 in any month.

**Figure 1.** *Evolution of Bluesky mentions of scientific publications from seven journals between January 2024 and March 2025*

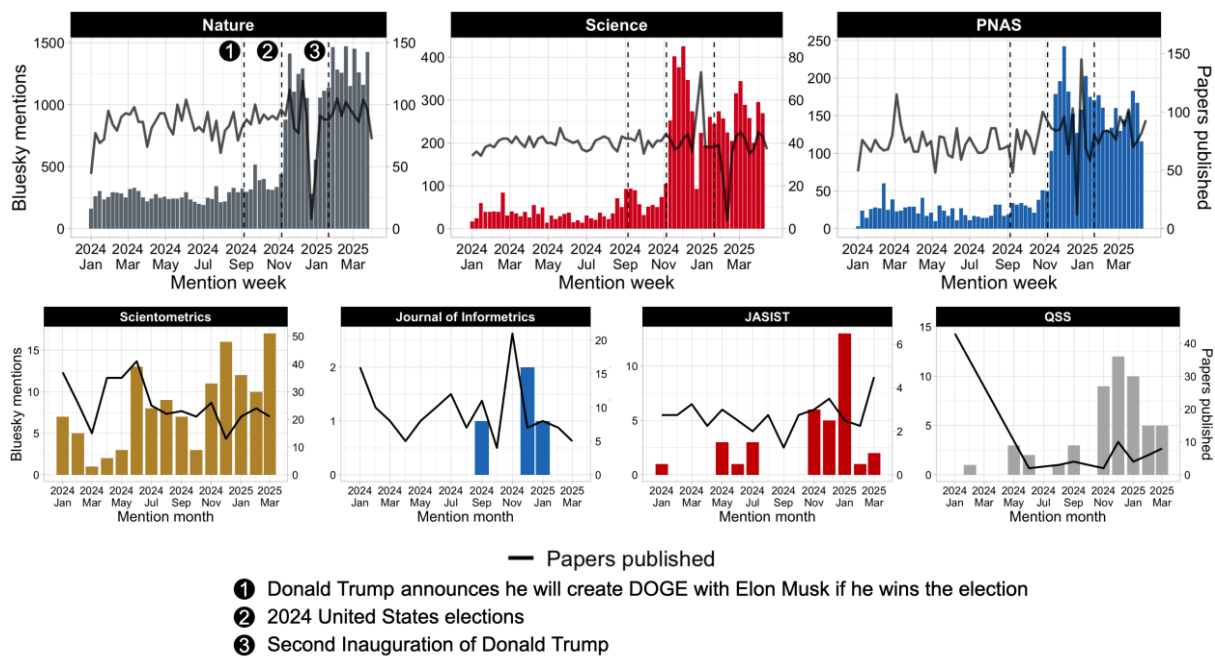

Throughout the period from January 2024 to March 2025, Bluesky has maintained a much smaller presence compared to X in terms of engaged users, meaning those who comment on papers, although its share has shown a clear upward trend as previously shown. This shift toward greater fragmentation has become especially evident since November 2024. As shown in Figure 2, while X dominated unique user mentions for all journals during the full period, typically holding over 80% in multidisciplinary journals and an even greater share in most LIS journals, the months following October 2024 reveal a much more balanced distribution between platforms. Bluesky's presence grows sharply, especially in the LIS field. In journals such as *JASIST* (65%) and *QSS* (56%), Bluesky even surpasses X as the main platform in unique user mentions, while in other cases, such as *PNAS* and *Journal of Informetrics*, the distribution between the two platforms becomes much more balanced. Taken together, these results point to a scholarly landscape where conversations are no longer concentrated on a single platform, but are now genuinely distributed between X and Bluesky.

**Figure 2.** Distribution of user mentions of papers on X and Bluesky by journal for papers published between January 2024 and March 2025, and between November 2024 and March 2025

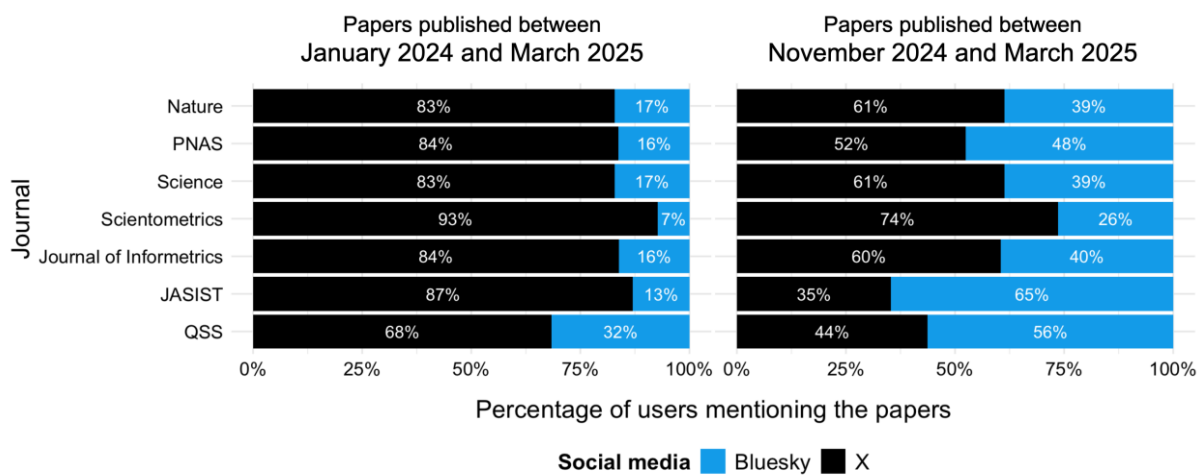

## 4. Discussion

Our exploratory analysis shows some evidence of user shift and increasing diversification between X and Bluesky, especially from late 2024 onwards, coinciding with major United States political events in which Elon Musk has actively intervened. This shift is especially visible in multidisciplinary journals such as *Nature*, where Bluesky's coverage nearly matches X, and user engagement increases notably. However, in LIS journals, Bluesky's presence is lower overall, but it is in the two United States journals (*JASIST* and *QSS*) where Bluesky even surpasses X in mentions, contrasting with the lower presence seen in the European journals (*Scientometrics* and *Journal of Informetrics*). This reflects that the response to platform changes is not only field-dependent, but also strongly shaped by political-geographical factors.

In this sense, user shift across platforms is a slow process which does not necessarily mean an abandonment of one of the platforms, favouring the other (Hou & Shiau, 2019), but tends to follow a push and pull model which could well end up with both platforms coexisting. This potential scenario could lead to further questions as to the already intricate question on the meaning of altmetric mentions (Robinson-Garcia et al., 2017), as we have two social media platforms which seem to be similar in functionalities but can potentially be hosting very distinct communities of users. Going back to the question on the moment in which Altmetric has decided to index Bluesky, it is certainly surprising as nothing of the sort happened with previous hypes such as the user shift to Mastodon two years ago (Chan, 2022). We can only speculate as to the reasons, one being the increasing opacity of X. However, this may need for re-investigation on the forming of communities discussing scientific literature (Arroyo-Machado et al., 2021) and adding complexity as there may be complementarities between both platforms.

In conclusion, evidence of a shift in user engagement between platforms is clear. However, it remains uncertain how effective this shift truly is, as it is still unclear how many users are maintaining parallel communication on both networks, whether they will abandon Bluesky to return to X or move to other platforms or even disengage from this space altogether. What is clear is that the altmetric hegemony of X may have come to an end, as for the first time there is a clear alternative in Bluesky, which even matches user engagement in ways that would have seemed unthinkable until recently. Only time will tell how effective and lasting this platform shift truly is.